\documentclass[aps,twocolumn,prd,showpacs,nofootinbib]{revtex4}
\usepackage{amsmath}
\usepackage{graphicx}
\usepackage{dcolumn}
\usepackage{bm}
\usepackage{amssymb}
\usepackage{latexsym}

\def\be{\begin{equation}}
\def\ee{\end{equation}}
\def\ba{\begin{eqnarray}}
\def\ea{\end{eqnarray}}

\begin{document}

\title{
On Perturbation Spectra of N-flation }

\author{Yun-Song Piao}
\affiliation{  
College of Physical Sciences,
Graduate School of Chinese Academy of Sciences, YuQuan Road 19A,
Beijing 100049, China}

\begin{abstract}

In this note we study the adiabatic perturbation spectrum of
N-flation with power law potential. We show that the scalar
spectrum of N-flation is generally redder than that of its
corresponding single field. The result obtained for that with
unequal massive fields is consistent with the recent numerical
investigation of Kim and Liddle.

\end{abstract}


\maketitle

The multi-field inflation is promising in the building of
inflation models, because it relaxes many bounds on single field
inflation models. Generally many fields can work cooperatively to
drive a period of inflation by assisted inflation mechanism
proposed by Liddle et.al \cite{LMS}, see also subsequent works
\cite{MW} and \cite{KO} for that with a spectrum of masses states,
even if any one of those field is not able separately to sustain
inflation\footnote{However, note that in Ref. \cite{AEGM}, a
realistic inflation model based on MSSM has been proposed.}. There
have been some interesting examples of multi-field inflation, see
Ref. \cite{Piao}, in which exponentially large number of fields
was required for a feasible theoretic realization of inflation.
Recently, Dimopoulos et.al \cite{DKMW} have shown that the many
axion fields predicted by string vacuum can be combined and lead
to a radiatively stable inflation, called N-flation, which may be
an attractive embedding of inflation in string theory.

In the simplest case where all fields have the same mass, there is
an attractor corresponding to the radial motion in field space.
Dimopoulos et.al assumed that all fields started with the same
initial condition, which equals to place them on this radial
trajectory. Thus they showed that the adiabatic perturbation has
the same spectral index as in the single field case. This has been
confirmed further in Ref. \cite{BW}. However, in the N-flation
setup there are large numbers of axion fields, and all fields have
different masses, which can be very densely spaced. Thus a
detailed exploration of the dynamics and the adiabatic
perturbation spectra in the unequal mass case seems indispensable.
The results were firstly mentioned in Ref. \cite{AL}. The detailed
study was made by Easther and McAllister \cite{EM}, however,
restricted to quite specific choices of initial conditions for the
fields. Recently, Kim and Liddle \cite{KL} have carried out
numerical investigations in which the random initial conditions
were used.
They found that the scalar spectral index has significant
dependence on the parameters of model, but when the number of
fields becomes enough large, the spectral index predicted will be
independent of initial conditions and enter into a long plateau.
In this note firstly we will educe a formula on the spectral index
of adiabatic perturbation of N-flation with power law potential,
and then with this formula we will try to give some
semi-analytical studies on the adiabatic perturbation spectrum of
N-flation with the unequal massive fields..

Though for a set of uncoupled fields with general power law
potential $\Lambda_i(\phi_i/\mu_i)^n$, where the subscript `i'
denotes the relevant quantities with the $i$th field and $n$ is
the same for all fields, there is no attractor solution
corresponding to the radial motion in field space, assisted
inflation remains. The reason is that the collection effect from
multi-field enhances the friction for individual field. Thus we
will use the slow-roll approximation for all formula in the
following. The efoldings number is \be N\simeq -{8\pi\over
m_p^2}\sum_i \int_{\phi_i}^{\phi_i^{e}} {V_i\over V_i^{\prime}}
d\phi_i \simeq {4\pi \over nm_p^2}\sum_i \phi_i^2 \,, \label{N}\ee
where the lower limits $\phi_i^e$ of the integrals correspond to
the end of inflation and have been neglected, which can be
reasonable for interesting cases.

The curvature perturbation of N-flation can be evaluated by using
Sasaki-Stewart formalism \cite{SS}, see also earlier Ref.
\cite{S}, in which the curvature perturbation on comoving slices
is expressed as the perturbation of efoldings number, which in
turn is given in terms of the inflaton perturbation on flat slices
after horizon crossing. For a set of above uncoupled fields,
the spectral index is given by \ba
n_{s}-1 &\simeq & -{m_p^2\over 8\pi}{\sum_i (V_i^{\prime})^2\over
V^2} -{m_p^2\over
4\pi \sum_i ({V_i\over V_i^{\prime}})^2} \nonumber\\
& + & {m_p^2\over 4\pi V}{\sum_i ({V_i\over V_i^{\prime}})^2
V_i^{\prime\prime} \over \sum_j({ V_j\over  V_j^{\prime}})^2 }
\nonumber\\ &= & -{1\over N}\left[1+{n\over 2}\cdot {\sum_i
({\Lambda_i \over \mu^n_i})^2 \phi_i^{2(n-1)}\sum_i \phi_i^2 \over
(\sum_i {\Lambda_i \over \mu_i^n}\phi_i^n )^2}\right],
\label{ns}\ea where
(\ref{N}) has been used in the last line and $V\equiv \sum_i V_i$.
This expression depends only on the quantities at horizon
crossing, however, it in fact is a good approximation, as was
discussed in Ref. \cite{KL}. Generally there also exist orthogonal
isocurvature perturbations \cite{L, GWBM}, see Ref. \cite{BTW} for
a review, which might or might not become important depending on
the evolution after inflation. However, here we will only focus on
the adiabatic perturbation for our purpose. Now we define
$g_i\equiv ({\Lambda_i\over \mu_i^n})/({\Lambda_1\over \mu_1^n})$
and $f_i\equiv \phi_i/\phi_1$, and then institute them into
(\ref{ns}). Thus $n_s$ can be written as \be n_s-1 \simeq -
{n+2\over  2N} \left[1 + {nR(g_i,f_i)\over n+2} \right]\,,
\label{nsr}\ee where \be R(g_i,f_i)= {\sum_{i< j} f_j^2f_i^2
\left( g_jf_j^{(n-2)}-g_if_i^{(n-2)}\right)^2 \over (\sum_i
g_if_i^{n})^2} \,.\label{rrgf}\ee This is our main result in this
note. The first term in the right side of Eq. (\ref{nsr}) is just
the result of single field inflation with power law potential,
while the second term can be regarded as a correction induced by
the variance between different fields, i.e. the differences
between their parameters and their initial conditions. The
interesting point of this formula is
that it can be seen very easily that the correction term is always
positive, which indicates that the scalar spectrum of N-flation is
generally redder than that of its corresponding single field.
Note also that this formula is invariance under the exchange
between $i$ and $j$, which means that the spectral index is
independent of the detailed array of serial number of fields.


For $n=2$, which corresponds to N-flation with massive fields
$m_i^2 \phi^2_i/2$, we can obtain from Eqs. (\ref{nsr}) and
(\ref{rrgf}) \be n_s-1\simeq -{2\over N}\left[1 + {R(g_i,f_i)\over
2} \right]\,,\label{ns2}\ee where \be
R(g_i,f_i)={\sum_{i<j}f_j^2f_i^2(g_j-g_i)^2\over \sum_i
(g_if_i^2)^2} \,,\label{n2}\ee where $g_i=m_i^2/m_1^2$.
Immediately we can see that when $g_j=g_i$, we have
$R(g_i,f_i)=0$, independent of the value of each field at the time
when the perturbation spectrum is calculated. Thus when the masses
of all fields are equal, the scalar spectrum of N-flation will be
the same as that of its corresponding single field, independent of
the initial conditions of fields.

The adiabatic perturbation spectrum of N-flation with unequal
massive fields is generally redder than that of its corresponding
single field has been mentioned in Ref. \cite{AL}, however, it
seems that it is not obvious to obtain such a conclusion and also
how the spectrum is dependent of unequal masses was not
illustrated. The relevant result was also obtained in Ref.
\cite{EM}, however, restricted to quite specific choices of
initial conditions and mass distribution for the fields. But in
our Eqs. (\ref{nsr}) and (\ref{rrgf}), one can straightly see that
the spectrum is redder than that of its corresponding single
field, not only for massive field but for general power law
potential, which is not dependent of the initial conditions and
the distribution of relative parameters, such as the masses and
couples of fields.

We discuss the perturbation spectrum of N-flation with unequal
massive fields with Eqs. (\ref{ns2}) and (\ref{n2}) in the
following. Firstly note that when there are only two massive
fields, the result of Lyth and Riotto \cite{LR} \be n_s-1\simeq
-{1\over N}\left[2+{f^2(g-1)^2\over \left(1+gf^2\right)^2}
\right]\ee can be obtained, where $g=m_2^2/m_1^2$ and
$f=\phi_2/\phi_1$. $n_s$ depends explicitly on $g$ and $f$, the
ratio of the values of two fields at the time with $N$ efoldings
number, This makes us hardly obtain the definite value of $n_s$.

For the number ${\cal N}\gg 1$ of fields, we take the mass
spectrum as $ m_i^2= m_1^2 \exp((i-1)/\sigma)$, where $i=1,2,...
{\cal N}$, as in Ref. \cite{KL}, and $\sigma$ gives the density of
fields per logarithmic mass interval. In the slow-roll regime the
field equation is $3h{\dot \phi_i}+m_i^2\phi_i \simeq 0$, thus the
fields obey the conditions \be f_i^2={\phi_i^2\over
\phi_1^2}\simeq\left({\phi_1\over
\phi_{1,0}}\right)^{2({m_i^2\over m_1^2}-1)}{\phi_{i,0}^2\over
\phi_{1,0}^2}\,,\label{fi2}\ee
where the subscript `0' denotes the initial value of field.

When ${\cal N}\ll \sigma$, the masses of all fields can be
approximately written as $m_i^2= m_1^2
\left[1+(i-1)/\sigma\right]$. Thus $g_i-1=(i-1)/\sigma\ll 1$,
which leads to \be g_j-g_i={j-i\over \sigma}<{{\cal N}\over
\sigma} \ll 1\ee and in the meantime since $g_i\simeq 1$,
Eq.(\ref{fi2}) can be reduced to $f_i^2\simeq \phi_{i,0}^2/
\phi_{1,0}^2$. These indicate that each term in numerator of
Eq.(\ref{n2}) is $\ll 1$ while each term in denominator is $\sim
1$, except there is a large hierarchy between $\phi_{i,0}$ and
$\phi_{1,0}$. Thus we can deduce $R(g_i,f_i)\ll 1$. Note that
adding the number of fields does not effect this conclusion, which
can be seen as follows. For a fixed large ${\cal N}$, the term
number of numerator is ${\cal N}({\cal N}-1)/2\sim {\cal N}^2 $,
thus the value of numerator is approximately ${\cal N}^2\delta$,
where $\delta \ll 1$, while the value of denominator is about
${\cal N}^2 {\cal O}(1)$. Therefore for the case that the number
of fields is far smaller that $\sigma$, the spectral index is
basically the same as that of single field. The main reason is
that in this case the increasing by degrees of mass of field is
negligible, which makes the result similar to that of N-flation
with same massive fields.

When ${\cal N}\simeq \sigma$, the difference of masses between
different fields begins to become important. Thus from Eq.
(\ref{n2}), the spectral index will shift towards the red
direction. The limit case is ${\cal N}\gg \sigma$. In this case
after some value $i_c$, we can have
$g_i=\exp{\left[(i_c-1)/\sigma\right]}\gg 1$. Note further that
generally $\phi_1/\phi_0 <1$. Thus from Eq.(\ref{fi2}), for an
enough large mass, we have $g_if_i^2\ll 1$, since the increasing
of $g_i$ leads to the exponential suppression of the value of
$f_i^2$. This suggests that after $i$ approaches some value $i_c$,
the contribution of $g_if_i^2$ to the denominator of (\ref{n2})
may be neglected and thus ${\cal N}-i_c$ terms after it, and
similar result may be applied for those of numerator. Due to
cut-off of the contributions from the fields with enough large
mass, the spread of spectrum towards red direction is certainly
not arbitrary large. Further it can be expected that the shift
generally constrained in a quite small region, as has been
numerically shown in Ref. \cite{KL}.

For $n= 4$, which corresponds to N-flation with
$\lambda_i\phi^4_i$ fields, we can obtain from Eqs. (\ref{nsr})
and (\ref{rrgf}) \be n_s-1\simeq -{3\over N}\left[1 +
{2R(g_i,f_i)\over 3} \right]\,,\ee where \be
R(g_i,f_i)={\sum_{i<j}f_j^2f_i^2\left(g_jf_j^2-g_if_i^2\right)^2\over
\sum_i \left(g_if_i^4\right)^2} \,,\label{n3}\ee where
$g_i=\lambda_i/\lambda_1$. The single field $\lambda\phi^4$
inflation has been ruled out by WMAP combined with SDSS
\cite{WMAP}, see also \cite{KKMR} and \cite{PE} for discussions on
the bounds of WMAP on the inflationary model space. Thus with more
fields, the spectrum will be generally more red, which is
certainly less interesting. However, it can be noted from
(\ref{n3}) that in the N-flation with $\lambda_i\phi^4_i$, even if
all couples $\lambda_i$ are equal, $R(g_i,f_i)$ also dose not
vanish, since the terms in the bracket of numerator are not only
dependent of $\lambda_i$ but the field value $\phi_i$, which is
distinctly different from that of $m_i^2\phi_i^2/2$. The cases for
$n>4$ are similar to that of $n=4$.

The tensor/scalar ratio $r$ is also an important inflation
quantity for observation, which as well as $n_s$ makes up of the
$r-n_s$ plane \cite{DKK}, in which different classes of inflation
modes are placed in different regions. In Ref. \cite{KL}, it has
been shown that the tensor/scalar ratio in N-flation model with
massive fields depends only on the efoldings number and is
independent of the number $\cal N$ of fields, their masses $m_i^2$
and initial conditions, and always has the same value as that of
its corresponding single field. In fact this result has been
noticed in Ref. \cite{AL}, and is also valid for the N-flation
with general power law potential $\Lambda_i (\phi_i/\mu_i)^n$
discussed here.

In summary, we educe the formula (\ref{nsr}) of the spectral index
of adiabatic perturbation of N-flation with power law potential.
This formula indicates that the scalar spectrum of N-flation is
generally redder than that of its corresponding single field. Then
with this formula we discuss the adiabatic perturbation spectrum
of N-flation with the unequal massive fields. We found that the
result is consistent with the numerical investigation of Kim and
Liddle \cite{KL}.

\textbf{Acknowledgments} The author would like to thank Richard
Easther, David Lyth, Anupam Mazumdar for valuable correspondences.
This work is supported in part by NNSFC under Grant No: 10405029,
90403032 and also in part by National Basic Research Program of
China under Grant No: 2003CB716300.

\end{document}